\documentclass[preprint,nofootinbib,aps,floatfix]{revtex4}
\usepackage{amssymb}
\usepackage{amsmath}
\usepackage{epsfig}

\newcommand{\beq}{\begin{equation}}
\newcommand{\eeq}{\end{equation}}
\newcommand{\beqa}{\begin{eqnarray}}
\newcommand{\eeqa}{\end{eqnarray}}

\newcommand{\lsim}{\mathrel{\rlap{\lower4pt\hbox{\hskip1pt$\sim$}}
    \raise1pt\hbox{$<$}}}         
\newcommand{\gsim}{\mathrel{\rlap{\lower4pt\hbox{\hskip1pt$\sim$}}
    \raise1pt\hbox{$>$}}}         

\newcommand{\GeV}{\mbox{\rm GeV}}
\newcommand{\TeV}{\mbox{\rm TeV}}

\def\vev#1{\langle#1\rangle}

\begin{document}

\vspace*{.5cm}

\title{Hiding the Higgs at the LHC}

\author{Omri Bahat--Treidel}\email{bahat@techunix.technion.ac.il}
\author{Yuval Grossman}\email{yuvalg@physics.technion.ac.il}
\author{Yoram Rozen}\email{rozen@techunix.technion.ac.il}
\affiliation{\vspace*{2mm}Department of Physics, Technion-Israel
  Institute of Technology, Technion City, Haifa 32000, Israel\vspace*{6mm}}



\begin{abstract}\vspace*{6mm}
We study a simple extension of the standard model where scalar
singlets that mix with the Higgs doublet are added.  This modification
to the standard model could have a significant impact on Higgs
searches at the LHC. The Higgs doublet is not a mass eigenstate and
therefore the expected nice peak of the standard model Higgs
disappears. We analyze this scenario finding the required properties
of the singlets in order to make the Higgs ``invisible'' at the
LHC. In some part of the parameter space even one singlet could make
the discovery of the SM Higgs problematic. In other parts, the Higgs
can be discovered even in the presence of many singlets.
\end{abstract}

\maketitle


\section{Introduction}
The Higgs particle of the Standard Model (SM) is expected to be
discovered at the LHC.  In extensions of the standard model, however,
the situation could be different. Modifications to the scalar sector
alter the experimental signatures of the Higgs boson in a model
dependent way. Therefore, there is no guarantee that a very general
Higgs boson can be found at the LHC.

The available experimental data provide constraints on the Higgs mass,
$m_H$ (for a review see \cite{b5}).  The strongest lower bound comes
from direct searches at LEP2, $m_{H}>114.4\;\GeV$ at 95\% CL
\cite{aaa}. An upper bound is derived from electroweak precision
measurements and reads $m_{H}<219\;\GeV$ at 95\% CL
\cite{2005em}. Since the sensitivity of electroweak precision
measurements to $m_H$ is logarithmic, we cannot exclude at a very high
confidence level the case where $m_H$ is just a factor of a few above
this limit.

One of the main goals of the LHC is to discover the Higgs boson. Both
the ATLAS and CMS collaborations will search for the Higgs boson in
the mass range of $10^{2}-10^{3}\;\GeV$. The Higgs is expected to be
discovered through different channels depending on its mass. In the
low mass regime, the most promising channel would be $H \to\gamma
\gamma$ \cite{b8}. For $m_H \gsim 150\;\GeV$, the preferred decay is
$H\rightarrow VV^{(*)}$ (with $V=Z,W$) with different substantial
decays of the vector bosons. These searches are expected to
provide at least a $5\sigma$ signal for the Higgs after few years of
operation of the LHC.

There are also several theoretical constraints on $m_H$ (see, for
example, \cite{b16}). For example, the unitarity bound reads
$m_{H}\lsim700\;\GeV$.  One can also consider the possibility that the
Higgs does not exist. This possibility gives rise to a constraint on
new physics scale $\Lambda\lsim 1\;\TeV$. Thus, we expect that the LHC
will find either the Higgs boson or some kind of new physics.

What if nothing is found at the LHC, that is, neither the Higgs boson
nor new physics? Such a scenario seems to imply that $(i)$ the Higgs
boson does not exist; $(ii)$ there is new physics that is responsible
for electroweak-symmetry breaking (EWSB); and $(iii)$ the experimental
signals of this new physics are such that it cannot be discovered at
the LHC. There is, however, another possibility: The Higgs exists and
it is responsible for EWSB but there is new physics that ``hides'' the
Higgs signals. Furthermore, this new physics does not show up in any
other channel and therefore cannot be discovered at the LHC.

Here we study such a scenario which hides the Higgs and does not show
any signal of new physics.  We extend the scalar sector of the SM by
introducing additional SM singlets which mix with the Higgs doublet of
the SM.  The resulting spectrum consists of many scalars. Each of
these scalars is mainly a singlet with a small component of the SM
doublet. Thus, the production rate for any of these mass eigenstates
is much smaller than that of a SM Higgs with the same mass. In the
limit of many singlets each mass eigenstate produces a very small
signal that cannot be separated from the background. In that case the
Higgs is practically hidden.

While our model is phenomenologically interesting, and demonstrate how
surprises may occur at the LHC due to simple new physics, it has no
clear theoretical motivation. In particular, while the model is not
necessarily fine-tuned, there is no reason why the new singlets have
weak scale masses and vevs. Our motivation is to study the
phenomenology of the model. Thus, at this stage, we do not try to find
a motivated fundamental framework that accommodates our model as its low
energy limit.

Several papers that study similar ideas has been published. A model
very similar to our was study by Espinosa and Gunion~\cite{Gunion:98}.
Like us, they study the effect of adding many singlets to the
SM. However, they focus on a $\sqrt{s}=500\;$Gev linear collider. They
found that such a multi singlet Higgs sector can be detected at such a
machine.

Singlet extensions of the SM where also discussed in other papers with
emphasis more on the effect of invisible decays of the Higgs.
Ref. \cite{Bij:march2006} discusses an $\cal O$(N) model without SSB
of the internal symmetry, and therefore there is no mixing with the
Higgs. In \cite{undewood:sep2006} a model with complex gauge singlet
was studied. The focus in that work had been put on the invisible
decay of the Higgs into the singlets goldstone modes.
Refs. \cite{Patt:2006fw,Calmet:2003uj} have studied a SM replica
called phantom sector. The phantom doublet acts as SM singlet which
might change the experimental signature. Two scenarios were examined.
The first, there is no mixing with the Higgs and the only effect is
the invisible decay (see also \cite{Burgess:2000yq}). The second, such
mixing does exist and the affect is reduced production rate.

Ref. \cite{Bij:april2006} explains an excess above LEP2 background
with large number of higher dimensional singlet scalar fields, which
mix with the SM Higgs. This model implies that none of the search
channels would work, while our analysis present a different
picture. In Ref. \cite{Dermisek:2005ar,Dermisek:2005gg} it has been
showed that the Higgs might have been missed at LEP2. This possibility
rises from the NMSSM, where the Higgs decays into two light $\cal CP $
odd Higgs bosons.

Ref.~\cite{O'Connell:2006wi} study a model similar to ours. Yet, while
we concentrate on the case of many singlets with large mixing with the
SM Higgs, \cite{O'Connell:2006wi} studies mainly the case of one
singlet with a small mixing angle.  In
Ref.~\cite{Manohar:2006gz,Manohar:2006ga} effective new operators were
introduced. These new operators can also hide the Higgs from being
discovered without showing any signal of new physics, but the
mechanism is different from ours.

Supersymmetric versions of our model study in
Ref.~\cite{Chang:2005ht,berger:nov2006,Barger:2006rd,Han:2004yd}
where signals of a supersymmetric model with one extra singlet
were investigated. Another supersymmetric model with a splitted
Higgs signal have been used to explain excesses of a Higgs signal
in LEP2~\cite{Dress:2005}. Our model is not supersymmetric and the
collider signatures are different than those discussed in these
papers.

\section{The model}

In order to understand the main features of our scenario we start with
a simple case where one singlet, $S(1,1)_0$, is added to the SM. For
simplicity we further introduce a $Z_2$ symmetry such that $S$ is odd
under it, while all other fields are even under this $Z_2$. Denoting
the SM Higgs doublet by $H$, the most general renormalizable scalar
potential is
\beq
\mu_H^2 |H|^2 + \dfrac{\mu_S^2}{2}  S^2 +
\lambda_H |H|^4 + \dfrac{\lambda_S}{4}S^4 +
\dfrac{\eta}{2} S^2|H|^2. \label{scalarV}
\eeq
In the following we assume that
\beq\label{our-ass}
\mu_H\sim\mu_S, \qquad
\lambda_H\sim\lambda_S\sim\eta.
\eeq
While our assumptions, that all dimensionful parameters are at the
same scale and all dimensionless couplings are of the same order, are
simple and not necessarily fine-tuned, they are not based on a
fundamental framework of new physics. We make them because they lead
to interesting phenomenology.

We are interested in the vacuum structure of this potential. Since the
Higgs vev is responsible for EWSB we demand $\vev{H} \ne 0$. As for
the vev of $S$, the solution $\vev{S}=0$ is not interesting as there
is no mixing between $S$ and $H$.  Thus, we consider only solutions
where $\vev{H} \ne 0$ and $\vev{S} \ne 0$. It is worth mentioning that
in general there is a large part of the parameter space where both
fields acquire a vev.

Next, we analyze the mass spectrum. We substitute
\begin{equation}
{\cal R}e(H)\to \dfrac{h+v_{H}}{\sqrt{2}}, \qquad
S \to {s}+v_S,  \label{6}
\end{equation}
where $h$ and $s$ are real scalar fields and $v_{H}$ and $v_{S}$ are
the vacuum expectation values of $H$ and $S$ respectively. The
mass-squared matrix in the $(h,s)$ basis is
\begin{equation}
M^{2}=
\begin{pmatrix}
\mu_H^{2}+3\lambda_H v_{H}^{2}+\frac{1}{2}\eta \,v_{S}^{2} & \eta
v_{H}v_{S} \\
\eta v_{H}v_{S} & \mu_S^{2}+3\lambda_S v_{S}^{2}+\frac{1}{2}\eta \,v_{H}^{2}
\end{pmatrix}
.  \label{7}
\end{equation}
Diagonalizing $M^2$, we get two mass eigenstates, $\phi_0$ and
$\phi_1$ with masses $m_0$ and $m_1$. We define $m_0\le m_1$ and due
to our assumptions we expect $m_0 \sim m_1$. We further consider only
cases where the two mass eigenstates are not close to be degenerate,
that is, $m_1-m_0 \gg \Gamma_0,\Gamma_1$. The two mass eigenstates are
related to the weak eigenstates $h$ and $s$ by a $2 \times 2$
orthogonal rotation matrix $V$
\beq
\begin{pmatrix} h  \\ s  \end{pmatrix}=V
\begin{pmatrix}\phi_0  \\ \phi_1\end{pmatrix},
\qquad
V=
\begin{pmatrix}
\cos \theta  & \sin \theta  \\
-\sin \theta  & \cos \theta
\end{pmatrix}.  \label{10}
\eeq
Note that $\theta$ can assume any value between $0$ and $\pi/2$.  In
general $\theta$ can be very small, but due to our assumption,
Eq. (\ref{our-ass}), we expect $\theta\sim O(1)$.  The model discussed
here contains five parameters. They can be chosen to be the five
parameters in (\ref{scalarV}). Instead, we can chose them to be the
two masses, $m_0$ and $m_1$, the two vevs, $v_{S}$ and $v_{H}$ and the
mixing angle $\theta$.

We are now in position to study the phenomenology of the model.
The couplings of the scalars to the SM fields can be obtained from
that of the SM Higgs by projecting onto the doublet component.  In
particular, we are interested in the coupling of a scalar to a pair of
SM fields, either fermions or vector bosons
\begin{equation}
\frac{V_{hi}}{v_{H}}\left(m_{f}\,\phi_i \bar{f}f +
m_Z^2\,\phi_i\, Z_{\mu }Z^{\mu} + 2 m_W ^2
\,\phi_i\, W_{\mu }^{+}W^{\mu -}\right).  \label{15}
\end{equation}
We see that the couplings are just the SM couplings projected by
$V_{hi}$.  The couplings between two scalars and two gauge bosons are
given by the SM ones multiplied by $V_{hi}V_{hj}$
\beq \label{15a}
\frac{V_{hi}V_{hj}}{2 v_H^2} \left(m_Z^2\,\phi_i \phi_j\, Z_{\mu }Z^{\mu}+
 2 m_{W}^{2}\, \phi_i \phi_j\,W_{\mu}^{+}W^{\mu -}\right).
\eeq
Last we need the self interactions term, i.e., interaction that
involve only scalars. The interesting part for our study is the
couplings that can be responsible for decays of a heavy scalar into
light scalars, $\phi_1 \to 2\phi_0$ and $\phi_1\to 3 \phi_0$. These
couplings are given by
\beqa &&
 \frac{1}{4}\Big[ \left(\lambda_S -\lambda_H -\left(\lambda_S
 +\lambda_H -\eta
\right) \cos 2\theta \right) \sin 2\theta \Big]\phi_1\phi_0^3
\label{13}+ \\ &&
\left[v_{\phi}\cos \theta \left(\left(3\lambda_S -\eta \right)
\sin^2\theta + \dfrac{\eta}{2}\cos^2\theta \right)
-v_{H}\sin \theta \left(\left(3\lambda_H -\eta \right) \cos^2\theta +
\dfrac{\eta}{2}\sin^2\theta \right) \right]\phi_1\phi_0^2 .
\notag
\eeqa
In general there are no specific relations between the strength of the
scalar couplings, Eq.~(\ref{13}), and the couplings between scalars
and gauge bosons, Eqs.~(\ref{15}) and (\ref{15a}). For example, the
coupling of $\phi_1\phi_0^2$ can be similar, smaller or larger to that
of $\phi_1W^+W^-$ .

We can generalize the above model by introducing $N$ new singlets,
$S_\alpha$, with $\alpha=1..N$. Again, we analyze the most interesting
case where all the scalar fields acquire vevs. The algebra is more
cumbersome, but we end up with a result similar to the case of one
extra singlet.  There are $N+1$ mass eigenstates $\phi_i$
($i=0..N$). We expand around the vacuum in a similar way as
Eq. (\ref{6}). In terms of the weak eigenstates, $\phi^W\equiv
(h,s_\alpha)$, the mass eigenstates $\phi$ are given by $\phi=V
\phi^W$, such that $V$ is an $(N+1)\times (N+1)$ orthogonal
matrix. The couplings to the SM fields are then given as in the one
singlet case by Eqs.~(\ref{15}) and (\ref{15a}). The analog of
Eq.~(\ref{13}) is more complicated. It can be obtained in a
straightforward way and we do not write it explicitly here. We only
mention that also in the more general case considered here the
couplings between the scalars can be smaller, similar, or larger with
respect to other couplings which involves gauge bosons.

\section{Phenomenology of the model}

Next we study the phenomenology of the $N$ singlets model. We first
look at the effect of this model on electroweak precision measurements
(see also \cite{undewood:sep2006}) and then move to discuss the
collider signatures.

The SM Higgs contribution to electroweak precision measurements comes
through the $S$ and $T$ parameters \cite{SandT}. That is, the gauge
boson self energies are the only numerically relevant diagrams with
the Higgs. Of course varying $m_H$ affects all observables, but in a
way consistent with changing just $S$ and $T$. Thus, in order to see
the effects of our model, all we need to do is to replace the SM Higgs
contributions to $S$ and $T$ with the sum of all contributions
weighted by the mixing angles. Consider a one-loop diagram with the
$i$th mass eigenstate. Its contribution to $S$ and $T$ is equal to
that of the corresponding SM diagram multiplied by $|V_{hi}|^2$. In
the leading log approximation, we therefore substitute
\beq
\log(m_h^2) \to \sum_i |V_{hi}|^2 \log(m_i^2).
\eeq
Thus, the bound on the Higgs mass in the SM is replaced by a bound on
a function of the masses and mixing angles. In particular, we can have
heavy mass eigenstates up to $1\;\TeV$ without violating the
electroweak data.

In order to discuss the implications of our model on collider searches
of the Higgs, we recall some issues regarding the search for the SM
Higgs. Depending on the Higgs mass, there are several decay channels
that are used to search for the Higgs. They are discussed at length in
Ref. \cite{b5} and are summarized in figs. 22 and 23 there. Roughly
speaking, we can say that
\begin{enumerate}
\item
Through most of the mass range, the Higgs is searched for by looking into
a resonance in different channels (like $H
\to \gamma\gamma$ or $H\rightarrow ZZ^{\left(
\ast \right) }\rightarrow 4\ell$).
\item
Around $m_h \gsim 170\;$GeV, where the dominant search channel is
$H\rightarrow WW^*\rightarrow l\nu l\nu$, and again at higher mass
($\gsim 400$ GeV), the Higgs is searched in a missing mass/momentum
channels such as $H\rightarrow ZZ\rightarrow \ell\ell\nu \nu$ and
$H\rightarrow WW\rightarrow \ell\nu jj$.
\end{enumerate}
A relevant point to the Higgs search is the width of the Higgs, $\Gamma_h$.  
The experimental resolution is expected to be $\sigma\sim 2\;$GeV
\cite{AtlasTDR} which is roughly the width of a Higgs with $m_h
\sim200\;$GeV.  For $\Gamma_h <\sigma$ a reduction of the Higgs width due to
added singlets
is practically impossible to detected, while for $\Gamma_h >\sigma$ this 
effect is more noticeable.

Now we move back to our model. The main effect of our model on
collider searches for the Higgs is that the cross section of each mass
eigenstate is suppressed compared to a SM Higgs of the same mass.  The
leading production process at the LHC is gluon fusion through one-loop
triangle diagram. Thus, the production cross section for each mass
eigenstate is suppressed by a factor of $|V_{hi}|^2$. In the limit of
many new singlets, $|V_{hi}|$ is small, and thus the cross section
become very small.

The other effects depend on the parameters of the model.  First
consider the scenario where decays of the form $\phi_i \to 2 \phi_j$ are
forbidden or negligible. Then, all the decay rates of the $i$th
mass eigenstate are suppressed by the same factor of
$|V_{hi}|^2$. Thus, the branching ratios are the same as those of a SM
Higgs with the same mass. The total width of each mass eigenstate is
smaller by a factor of $|V_{hi}|^2$ compared to the width of a SM
Higgs with the same mass.

In the case of a resonance search in the above scenario, the Higgs
width affect our model.  At low masses the Higgs width is small
compared to the experimental resolution.  Then the signal of each mass
eigenstates is reduced by $|V_{hi}|^2$. (The width is also reduced by
the same amount but this reduction cannot be noticed.)  With many
singlets, when $|V_{hi}|^2$ is very small for all $i$, the signal
significance will drop below detection level.

At Higher masses, when the width of each mass eigenstate, $\Gamma_i$, is large,
$\Gamma_i > \sigma$ the division of the signal between the singlets
reduces the significance of each resonance by $|V_{hi}|$. The reason
is that while the total signal is reduced by $|V_{hi}|^2$, this
reduction simultaneously affects the width of the resonance. 

In the case of a non-resonance search, the mass eigenstates contribute
to the missing energy signal. Hence, the combined excess of these
eigenstates over the background will be similar to that of a SM Higgs
with $m_h \gsim 400\;$GeV. In this case it is possible to hide the
Higgs signal by adding light mass eigenstates in the resonance search
mass range.

Last we discuss the scenario where decays like $\phi_i \to 2 \phi_j$ are
important. In particular, the interesting case is when all the heavy
scalars decay almost entirely to the lightest one. In that case the
situation is similar to the SM Higgs. Only one mass eigenstate is
produced and its branching ratios are the same as a SM Higgs with the
same mass. Yet, the production cross section and width are smaller
than for a SM Higgs. This is because the production cross section for
a heavy mass eigenstate is always less than half that of the light
one. Thus, the fact that a heavy mass eigenstate decays into two light
scalars cannot compensate for the reduction in the production rate 
and the parameter space allow for the possibility of the Higgs being hidden.

\section{Examples}

In the following we work out a few examples showing how additional
singlets can hide the Higgs signal at the LHC. These examples are all
within the suppressed $\phi_i \to 2 \phi_j$ scenario. In the first
example we deal with a failure of a specific decay channel namely the
$H\rightarrow ZZ\rightarrow 4\ell$.  The second example discusses the
case of a missing mass channels such as the $WW^*\rightarrow \ell\nu
\ell\nu$. In both examples we assume that the additional singlet masses
are all in the range best suited for discovery in the discussed
channel. Finally we give a third example which is the minimal solution
for the LHC. In this example we follow the title of our paper by
adding the minimal number of singlets needed to hide the Higgs at the
LHC regardless of the search channel.

\begin{figure}[t]
\epsfig{figure=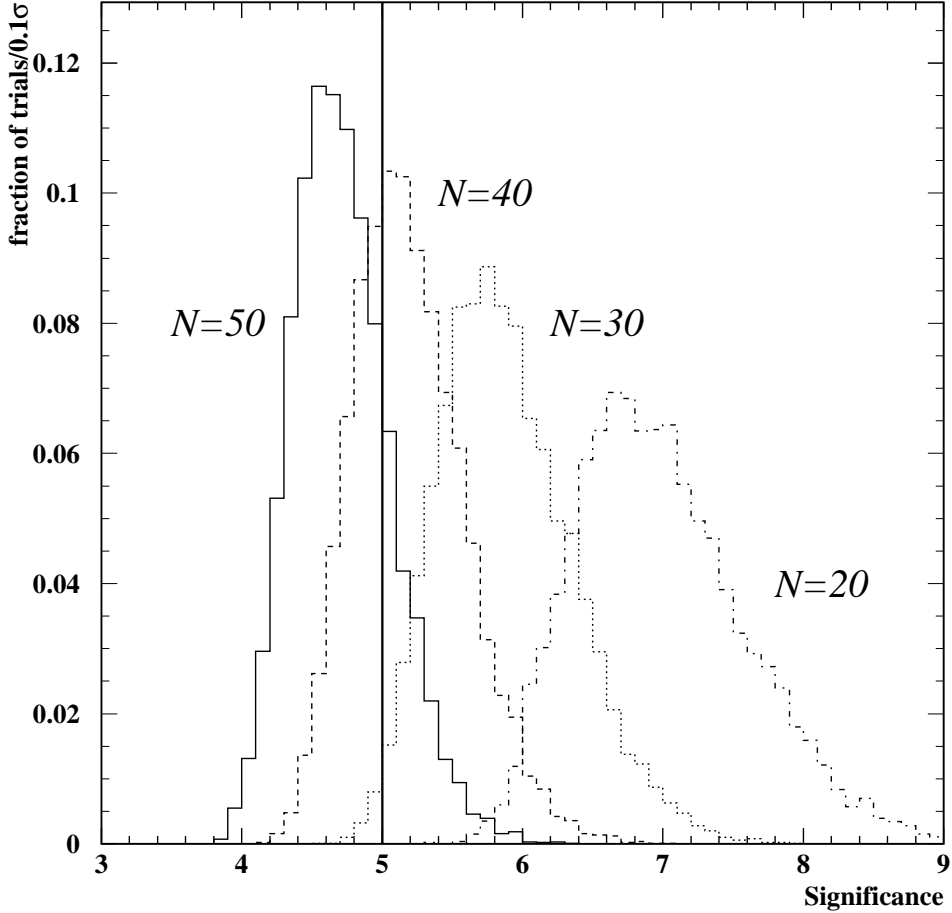, width = 14cm}
\caption{Significance of the most significant singlet in 
20-50 singlets toy Monte Carlo experiments. The vertical line at
$\sigma=5$ represents the discovery threshold. See text for details.}
\label{fig1}
\end{figure}

\subsection{$H\rightarrow ZZ\rightarrow 4\ell$}
A search for the Higgs in this
channel is most effective in the mass range 180-700 GeV. It is often
dubbed the ``golden channel'' due to the expected high signal
significance. If the Higgs mass is about 300 GeV it is expected to
yield a very high signal significance already in the first year of
operation with as little as 10 fb$^{-1}$ of integrated
luminosity. Failure to discover the Higgs with three times that
luminosity seems unimaginable assuming a SM Higgs. However, if one
adds 12 singlets in the mass region 200-300 GeV with mixing constants
$|V_{hi}|^2=0.03$ for $i=1..12$, additional 10 in the next 150 GeV
with $|V_{hi}|^2=0.04$ for $i=13..22$, and finally one more at around
600 GeV with $|V_{h23}|^2=0.24$ the signal significance drops below
four at any given mass. This potential distribution of singlet masses
was chosen with the Higgs width and experimental resolution in
mind. However, while the mass spectrum will differ significantly from
the expected SM Higgs signal it will also differ from the expected
background shape and yield, indicating some ``other'' source. Figure
\ref{fig1} shows the significance distribution of the most significant
singlet in a 50 (40,30,20) singlets experiment randomly drawn. The
significance is calculated for a 30 fb$^{-1}$ of integrated luminosity
for signal and background of the decay $H\rightarrow ZZ\rightarrow 4l$
in the mass range 180-420 GeV. It can be seen that already with 30
arbitrary drawn singlets in that mass range, part of the simulated
experiments have no singlet with more than 5$\sigma$ significance. The
mixing constants, $|V_{hi}|^2$, were drawn from a Gaussian
distribution whose mean was set to 1/(number of singlets) and a width
of half that number. If one were to include the higher mass range for
this channel (420-700 GeV) and allow for a few singlets to occupy that
region where the expected significance is lower, a larger fraction of
the shown distributions will be found below 5$\sigma$ which is the
case given in the above example. This is since the lower significance
allow for larger value of $|V_{hi}|^2$ for the singlets in the high
mass range.

\subsection{$H\rightarrow WW^*\rightarrow l\nu l\nu$}
If the mass of all the
 singlets is in the vicinity of 170 GeV where the dominant decay channel is 
$H\rightarrow WW^*\rightarrow l\nu l\nu$ the resulting signal will 
differ only slightly from the expected SM Higgs signal. This is due to the 
inability to fully reconstruct the Higgs mass. Hence the signal will be 
observed as an excess of event over the expected background. No possibility to 
hide the Higgs in this region if one insists on a solution of singlets solely 
in this mass range.

\subsection{Hiding the Higgs at the LHC}
Regardless of the specific examples
above, the minimal solution in our model for hiding the Higgs for 100
fb$^{-1}$ will be with three singlets at about 118, 124 and 130 GeV
and about equal value for the three $|V_{hi}|^2$. In which case none
of the mass eigenstates will be discovered and the overall number of
observed events will be consistent with the background hypothesis.


\section{Discussion and conclusions}

The Higgs boson is expected to be discovered at the LHC. Depending on
its mass, different channels will be used to discover it. The standard
model will be in a very bad position if the Higgs is not found. In
this work we have shown that additional singlets might explain an
absence of a Higgs signal without any signal of new physics. We
analyzed scenarios corresponding to different masses in the range of
$10^2 \lsim m_{i}\lsim 10^3\;$GeV. We assumed that all dimension-full
parameters are of the order of the weak scale and all dimensionless
parameters are of order one. In particular we asked how many singlets
are needed in order to ``hide'' the Higgs. The answer depends
crucially on the model parameters.  In some cases, in particular when
the mass eigenstates are close to $100\;$GeV, we found that as little as 2-3
singlets could reduce the significance below discovery
level.  In other cases, mainly when many of the masses are roughly
above $300\;$GeV we found that tens of singlets are needed to hide the
Higgs.  


We have concentrated on the Higgs search at the LHC. In fact, it could
affect the searches for the Higgs also at LEP and the Tevatron and it is possible that
the Higgs signal is hidden by a many-singlet solution. Yet, we did not
investigate this issue in details. For the case of one extra singlet
such a study was done in \cite{O'Connell:2006wi}.

To conclude, we present a model in which the standard model Higgs
field generates electroweak symmetry breaking but still the Higgs
particle cannot be discovered at the LHC. Our model is very simple,
and while it is not based on a well motivated theoretical framework,
it serves as an example that the SM Higgs mechanism can escape
detection at the LHC.


\section*{Acknowledgments}
We are grateful to Ehud Duchovni, Andrey Katz, Amos Ori, and Witold
Skiba for useful discussions and Yossi Nir, Yael Shadmi and Ze'ev
Surujon for comments on the manuscript. The work of Y.G. is supported
in part by the Israel Science Foundation under Grant No. 378/05.  The
work of Y.R. is supported in part by the Israel Science Foundation
under Grant No. 1446/05.


\end{document}